\documentstyle[times,pramana,epsf,floats]{ias}
\begin{document}
\mark{{Photon multiplicity Measurements...}{Bedangadas Mohanty}}

\title{Photon Multiplicity Measurements : From SPS to RHIC and LHC}
\author{Bedangadas Mohanty}
\address{Institute of Physics, Bhunaneswar 751005, India \\
         e-mail : bedanga@iopb.res.in}



\abstract{
  Results from the photon multiplicity measurements 
  using a fine granularity preshower photon multiplicity detector 
  (PMD) at CERN SPS are discussed. These include study of
  pseudo-rapidity distributions of photons, scaling of photon
  multiplicity with number of participating nucleons, 
  centrality dependence of $<p_{T}>$ of photons, 
  event-by-event fluctuations in photon multiplicity and 
  localised charged-neutral fluctuations. Basic features of the PMD
  to be used in STAR experiment at RHIC and in ALICE experiment at LHC 
  are also discussed. 
}

\maketitle

\section{Introduction}

  Measurement of photon multiplicity in relativistic heavy-ion
  collisions is complimentary to the well established methods
  of charged hadron measurements and it shows a great promise
  in studying the various aspects of the reaction mechanism
  of phase transition from hadronic matter to Quark-Gluon Plasma
  and dynamics of particle production.
  In heavy-ion collisions it is important to correlate information
  obtained from various global observables (such as charged
  particle multiplicity, mean transverse momentum ($<p_{T}>$)
  and transverse energy ($E_T$) ) for proper understanding of the
  physics processes occurring in the reaction. Photon multiplicity 
  is an additional global observable.  More specifically photon  
  multiplicity provides an unique opportunity to study the changes 
  in the relative population of the electromagnetic and hadronic 
  components of the multi-particle final state. These and several 
  other aspects are discussed in this paper.

  The paper is organised as follows, first we give a brief description 
  of the photon multiplicity detectors (PMD) used at SPS
  experiments. In section~3 we discuss the various physics
  issues addressed by the PMD at SPS, such as the study of pseudo-rapidity
  distribution of photons ($dN_{\gamma}/{d\eta}$), 
  scaling of total number of photons with
  number of participating nucleons, measurement of event-by-event mean 
  transverse momentum ($<p_{T}>$) of photons, 
  event-by-event fluctuation in photon 
  multiplicity and search for domains of disoriented chiral
  condensates (DCC) through the study of localised charged-neutral
  fluctuations. A comparison to the results from charged
  particles will be made wherever possible. The study of 
  azimuthal anisotropy of photons is covered in an separate article in 
  this volume. In section~4 we discuss the PMD to be installed in STAR 
  experiment at RHIC and ALICE experiment at LHC. 
  This is followed by a summary in the last section.

\section{Photon multiplicity detectors at CERN SPS}

\begin{figure}
\setlength{\unitlength}{1mm}
\begin{picture}(130,50)
\put(10,-2){
\epsfxsize=6.1cm
\epsfbox{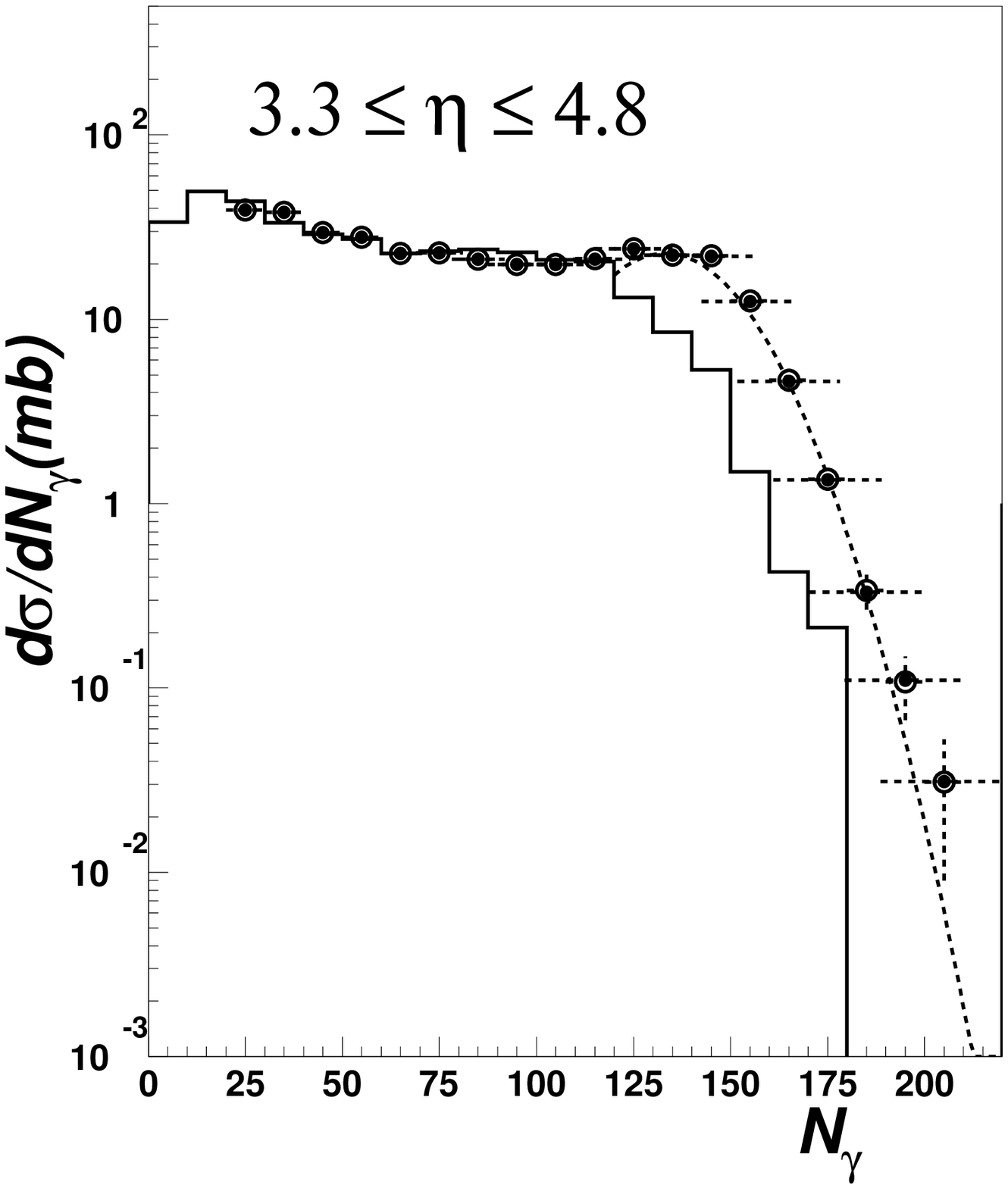}
}
\put(70,-6.0){
\epsfxsize=7.1cm
\epsfysize=7.1cm
\epsfbox{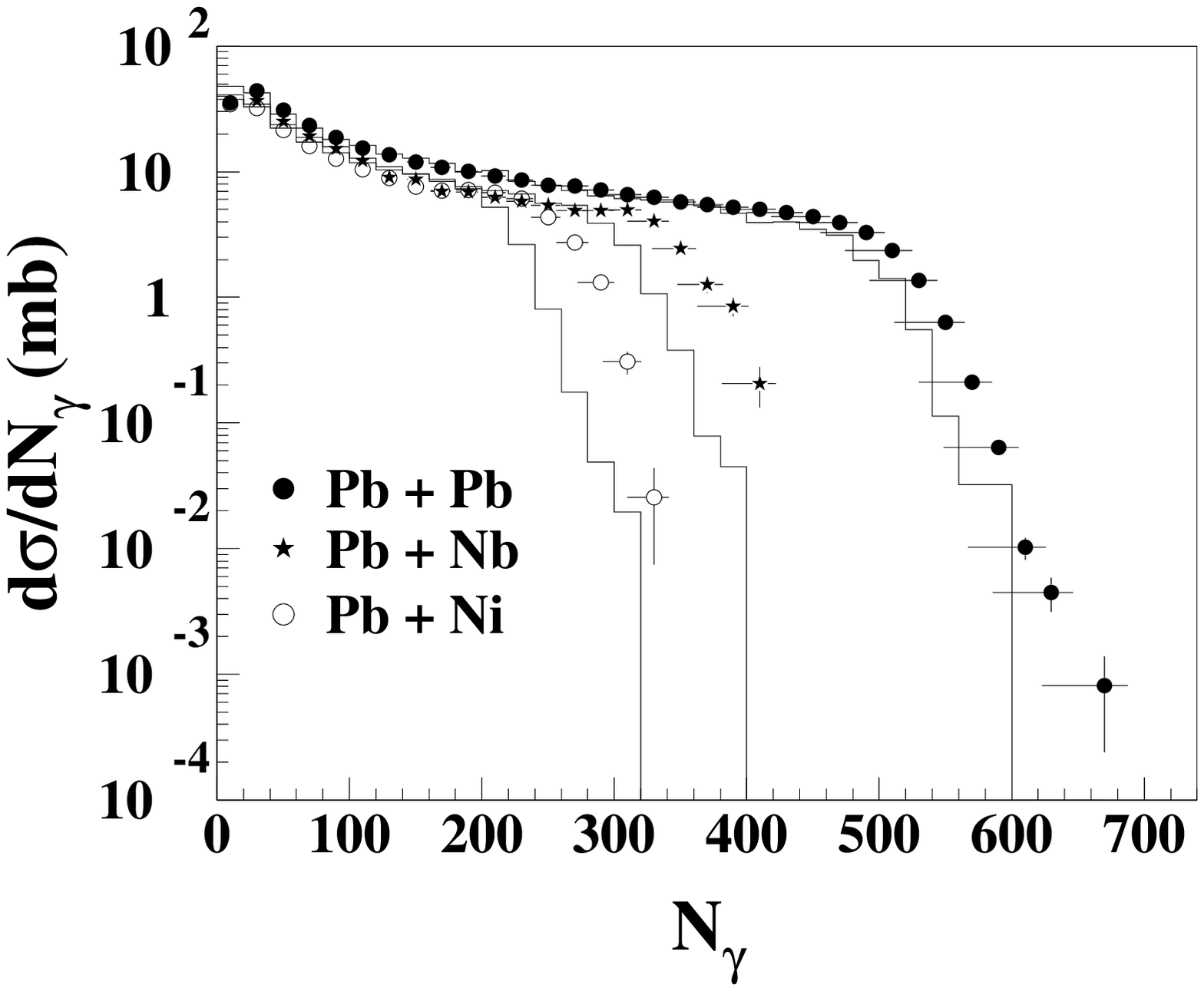}
}
\end{picture}
\vspace*{0.4cm}
\caption{
Minimum bias inclusive photon cross sections for S+Au reaction at 
200$\cdot$AGeV (left plot) and for Pb+Ni, Pb+Nb, and Pb+Pb reactions
at 158$\cdot$AGeV (right plot). Solid histograms are the
corresponding distributions obtained from the VENUS event generator.
}
\label{minbias}
\end{figure}

A fine granularity preshower PMD was implemented in the WA93
experiment at CERN SPS~\cite{wa93_pmd}, 
allowing the $first~ever$ measurement of
multiplicity, rapidity and azimuthal distributions of photons in 
ultra-relativistic heavy-ion collisions. The basic features of the detector 
are given in Table~1. The minimum bias distribution of the photon
multiplicity as measured by the PMD for S + Au at 200 AGeV 
in WA93 experiment is shown in Figure~\ref{minbias}. 
The distribution has been obtained for the 
full azimuthal coverage of the PMD and has been compared to results
obtained from the VENUS event generator. It is observed that VENUS
under predicts photon multiplicity for central events.  Subsequently
a similar PMD was also installed in the WA98 experiment~\cite{wa98_pmd}. 
The basic
features of the PMD in WA98 experiment are also given in Table~1. The
minimum bias distribution of photon multiplicity as measured by the
PMD in WA98 experiment is shown in Figure~\ref{minbias}. The
experimental results for different target ions has been compared to
those obtained from the VENUS event generator. VENUS is found to under 
predict photon multiplicity for central collisions, and it under
predicts more for asymmetric ion collisions (Pb + Nb and Pb + Ni).

\begin{table}[h]
\caption{PMD in WA93 and WA98 experiments}
\begin{tabular}{lll}
Basic features & WA93 & WA98 \\ \hline
Data taking & 1991 and 1992 & 1994-1996 \\ 
Beam and target & S + Au \& S + S & Pb + Pb, Pb + Ni \& Pb + Nb \\ 
Energy & 200 A GeV & 158 A GeV\\ 
No. of scintillator pads & 7600 & 53000 \\
Readout & CCD camera & CCD camera \\
Distance from target & 10.09 m & 21.5 m \\ 
$\eta$ coverage & 2.8 - 5.2 & 2.5 - 4.2 \\
$\eta$ coverage with full $\phi$ & 3.3 - 4.8 & 3.2 - 4.0 \\
$p_{T}$ acceptance & $\ge$ 20 MeV/c & $\ge$ 30 MeV/c \\ 
Photon counting Efficiency (central to peripheral) & 65 - 75\% & 68 - 73 \%\\ 
Purity of photon sample(central to peripheral) & 70\% & 65 - 54 \%\\ 
\end{tabular}
\end{table}

Both WA93 and WA98 experiments include several other detectors for 
charged particle multiplicity and momentum measurements along with 
trigger detectors for defining the centrality of the reaction. The 
common coverage of the charged particle multiplicity detector and the 
calorimeter (for measuring $E_{T}$) with PMD was used for
event-by-event measurement of $<p_{T}>$ of photons and 
search for possible formation of DCC.

\section{Physics issues addressed by PMD at SPS}

In this section we discuss the various physics issues addressed
by the PMD at CERN SPS. These includes study of
$dN_{\gamma}/{d\eta}$, scaling of photons,
event-by-event $<p_{T}>$ of photons,
event-by-event fluctuation in photon multiplicity and formation of DCC.

\subsection{Pseudo-rapidity distribution of photons}

\begin{figure}
\setlength{\unitlength}{1.1mm}
\begin{picture}(110,50)
\put(10,5){
\epsfxsize=7.1cm
\epsfysize=7.8cm
\epsfbox{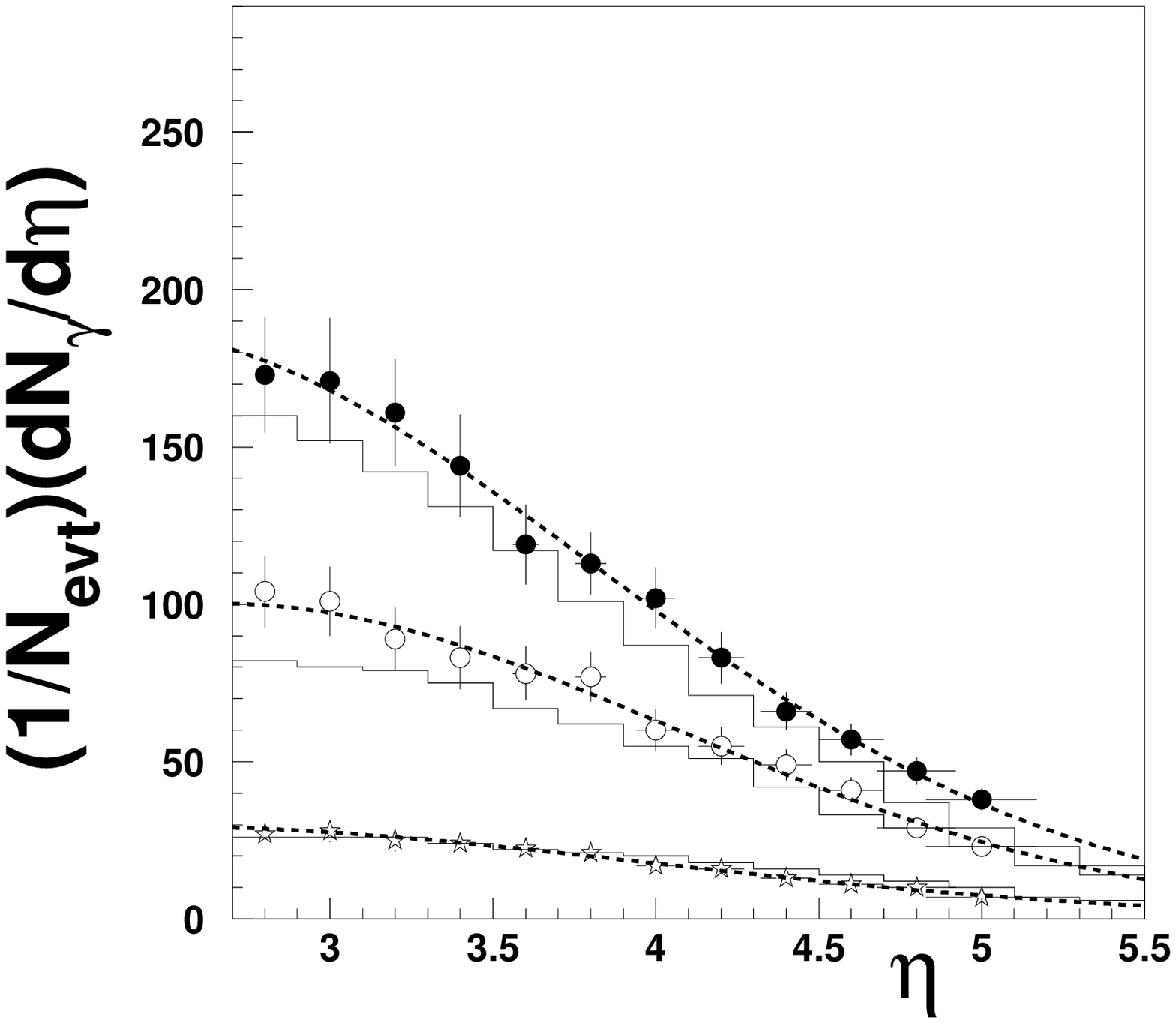}
}
\put(70,5){
\epsfxsize=7.1cm
\epsfbox{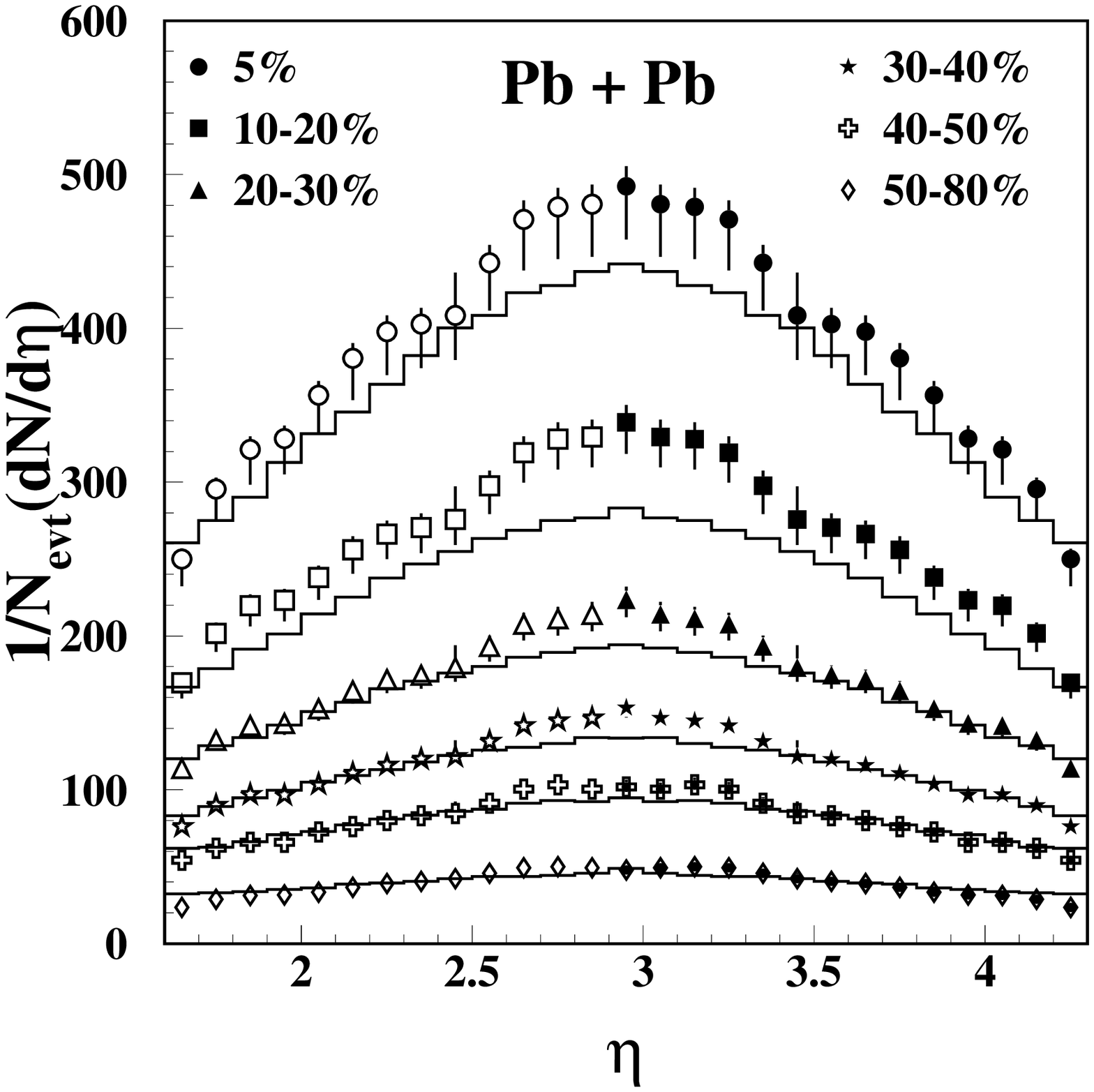}
}
\end{picture}
%\vspace*{-1cm}
\caption{
Pseudo-rapidity distributions of photons for S+Au reaction at
200$\cdot$AGeV (left plot) and Pb+Pb reaction at 158$\cdot$AGeV
(right plot). The solid histograms are the corresponding distributions
obtained from  the VENUS event generator.
}
\label{rap_dist}
\end{figure}

One of the challenges in relativistic heavy-ion collisions is the
large number of particles produced. Measurement of 
particle density in rapidity
is a convenient way to describe heavy-ion collisions. It has also
been suggested that fluctuations in pseudo-rapidity distributions
is a signature of phase transition from hadronic matter to Quark-Gluon
plasma.  Further, pseudo-rapidity density can be
related to a thermodynamic quantity,entropy density, in heavy-ion
collisions. All these motivate 
us to study the pseudo-rapidity distributions of photons (
$dN_{\gamma}/{d\eta}$). We have studied
$dN_{\gamma}/{d\eta}$ in WA93 and WA98
experiments at CERN SPS~\cite{rap_wa93pmd,rap_wa98pmd}. 
These distributions for S + Au and Pb + Pb
reactions for different centrality classes are shown in  
Figure~\ref{rap_dist}. They are
found to be good gaussians. The results from VENUS are also shown in
the figures. One observes that VENUS under predicts data for central
collisions.

\begin{figure}
\setlength{\unitlength}{1.1mm}
\begin{picture}(110,50)
%\vspace*{1.0cm}
\put(10,5){
\epsfxsize=7.5cm
\epsfbox{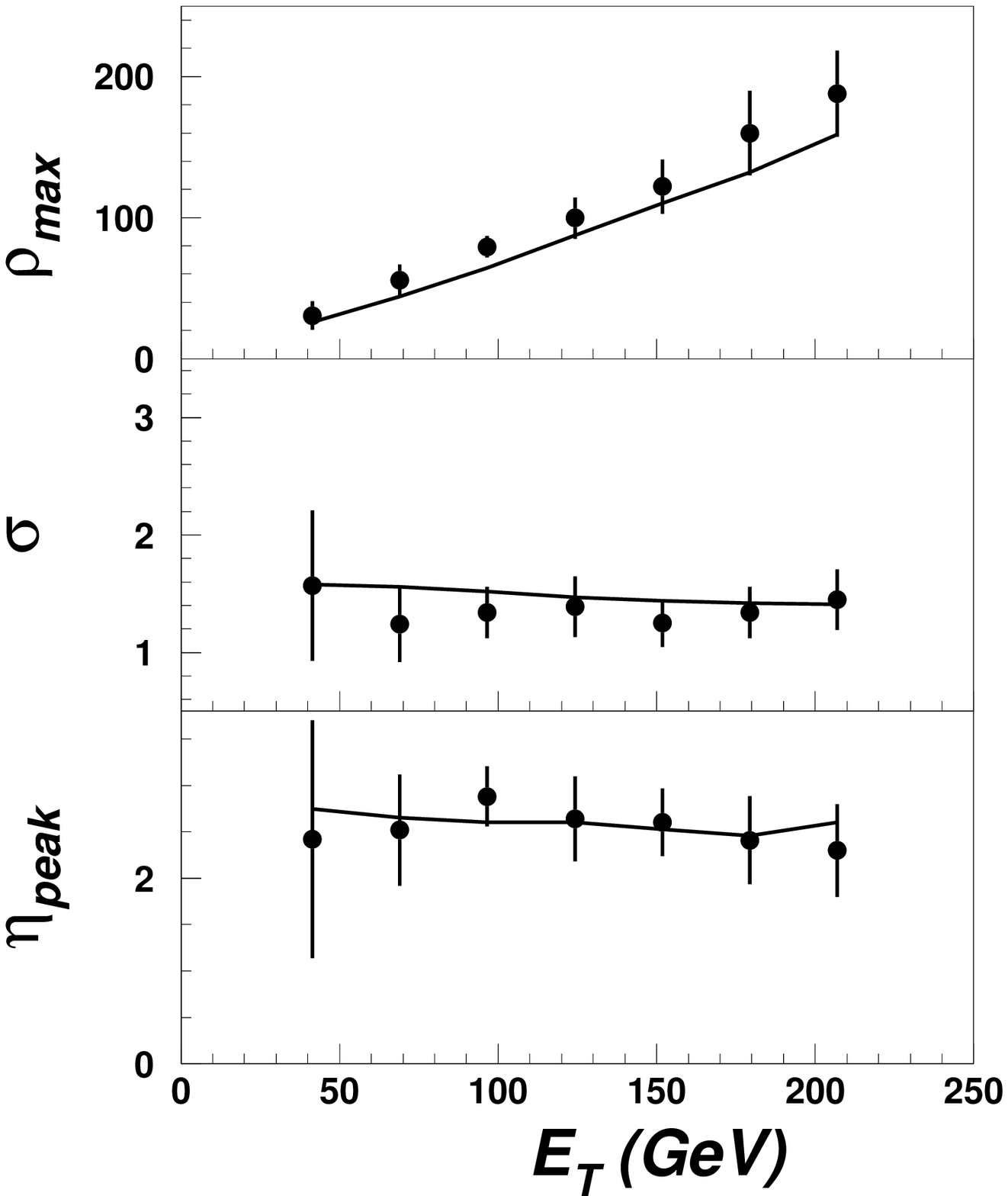}
}
\put(70,5){
\epsfxsize=7.5cm
\epsfbox{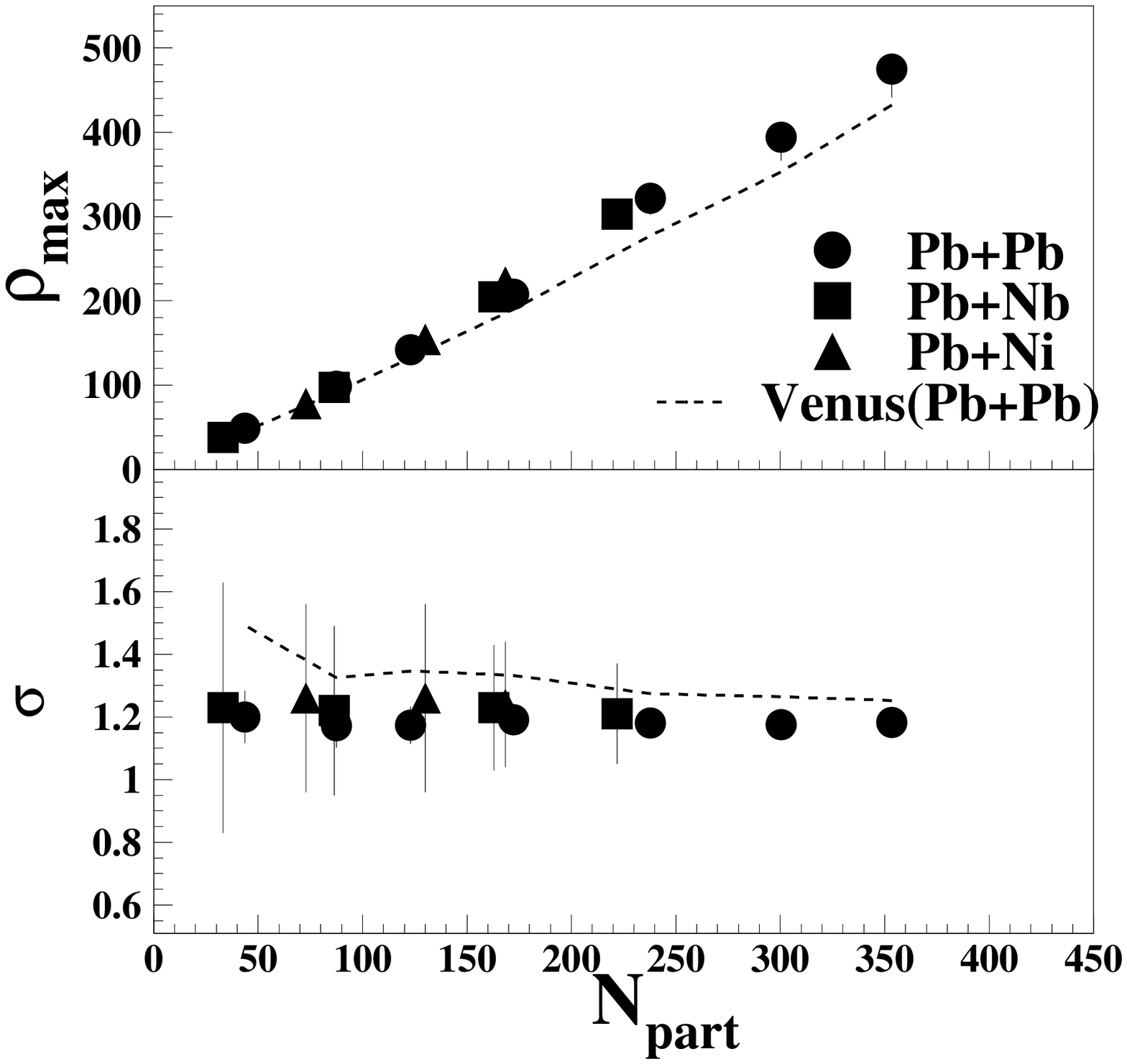}
}
\end{picture}
%\vspace*{-1cm}
\caption{
Pseudo-rapidity density ($\rho_{max}$), and width of the pseudo-rapidity 
distribution ($\sigma$) are shown as a function of centrality of the
reaction (defined through the $E_T$
values for WA93 results (left plot) 
and number of participating nucleons for WA98
results (right plot)). 
}
\label{shape}
\end{figure}

The best way to study the pseudo-rapidity distributions is to 
look at the shape parameters of the distributions. The shape
parameters are the pseudo-rapidity density at mid-rapidity
($\rho_{max}$), width of the pseudo-rapidity distribution ($\sigma$)
and pseudo-rapidity peak ($\eta_{peak}$). These are shown in
Figure~\ref{shape} as a function
of transverse energy and number of participating nucleons for WA93 and
WA98 experiments respectively.
The results have been compared to those obtained from VENUS event
generator. The pseudo-rapidity density at mid-rapidity is found to
increase with increase in centrality, which can be understood from
simple geometrical picture of the collision. Results from VENUS is
also found to follow the similar trend. For higher centrality of
the reaction VENUS under predicts the data. The width of the
pseudo-rapidity distributions for various centrality classes are found to 
be similar within the quoted errors and the trend is well explained by
results from VENUS. The pseudo-rapidity peak for the WA98 experiment
was found to be $2.92$.

\subsection{Scaling of photons}

\begin{figure}[htbp]
\epsfxsize=8.0cm
\centerline{\epsfbox{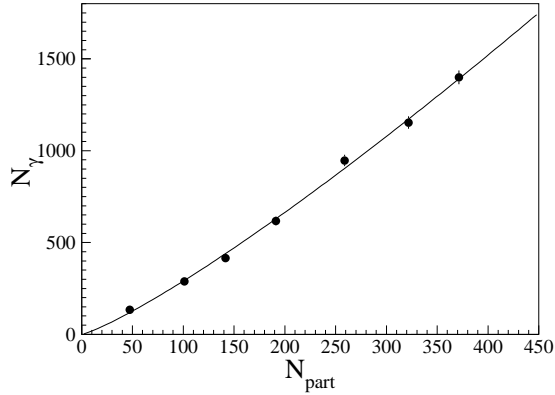}}
\caption{ Scaling behavior of photons. Integrated number of photons
  are plotted as a function of number of participating nucleons. The
  solid lines show power-law fit to the data, which yields the value
  of exponent, $\alpha$ = 1.12 $\pm$ 0.03. }
\label{scaling}
\end{figure}

It is important to study the scaling of particle multiplicity 
as they test the various models for particle
production. Also various experimental signatures require comparison of
observables of different system sizes, hence a proper understanding of 
scaling is essential. While scaling with number of collisions arises
naturally in a picture of a superposition of nucleon-nucleon
collisions, with possible modifications by initial state effects, the
participant scaling is more naturally related to a system with strong
final state re-scattering, where the incoming particles lose their
memory and every participant contributes a similar amount of
energy. The scaling behavior can therefore carry important
information on the reaction dynamics. 
It is therefore of interest to study the scaling properties
with respect to number of participants or collisions.
We have studied the scaling of total number of photons with number of
participating nucleons~\cite{rap_wa98pmd} . 
The results for photons from Pb+Pb collisions at 
158$\cdot$AGeV are shown 
in Figure~\ref{scaling}. Fitting the data points to the function
$C~\times~N^{\alpha} _{part}$, yields the value of $\alpha$ to be
$1.12~ \pm ~0.03$. Similar analysis for charged particles in the same
experiment and using the data from the silicon pad multiplicity
detector gives a value of $\alpha$ = 1.07 $\pm$
0.05~\cite{scaling_ch}. Within the
quoted systematic errors the value of $\alpha$ are similar for both
photons and charged particles and
indicates a deviation from the picture of a 
naive Wounded Nucleon Model ($\alpha$ = 1).

\subsection{Event-by-event mean transverse momentum of photons}

Variation of $<p_{T}>$ of produced particles with global
multiplicity could provide signature of phase transition from hadronic 
matter to Quark-Gluon plasma. The expected behavior will be similar
to that of variation of entropy density with temperature, where
temperature increases with increase in entropy density before phase
transition then remains unchanged during the transition and finally
again increases with entropy density after phase transition. Entropy
density is related to pseudo-rapidity density and transverse momentum
to temperature. Hence a study of variation of transverse momentum with
pseudo-rapidity density (or centrality of reaction) can be an 
experimental signature of QGP formation.

\begin{figure}
\setlength{\unitlength}{1.1mm}
\vspace*{1.5cm}
\begin{picture}(130,50)
\put(10,5){
\epsfxsize=7.5cm
\epsfysize=7.5cm
\epsfbox{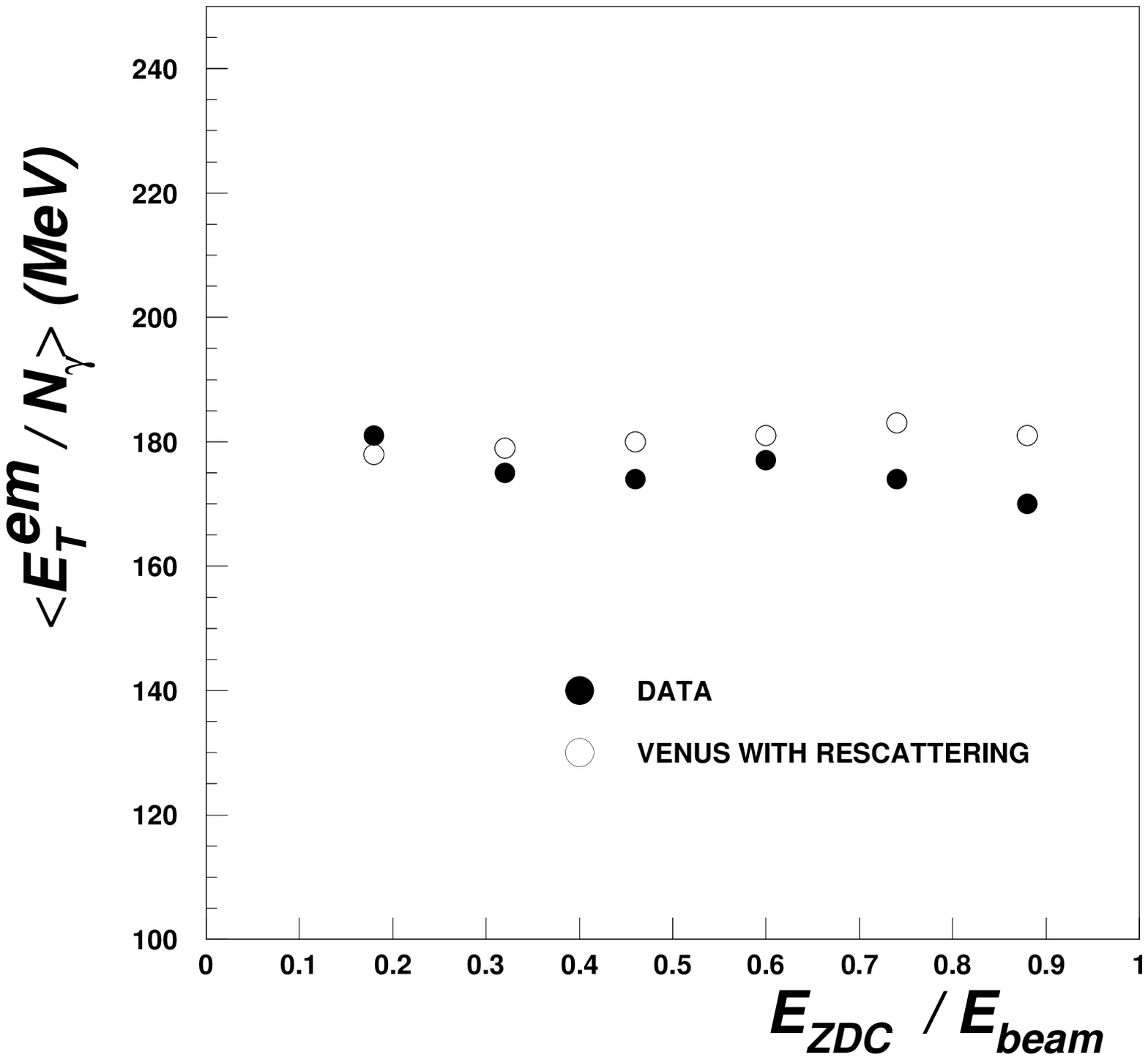}
}
\put(70,5){
\epsfxsize=7.0cm
\epsfbox{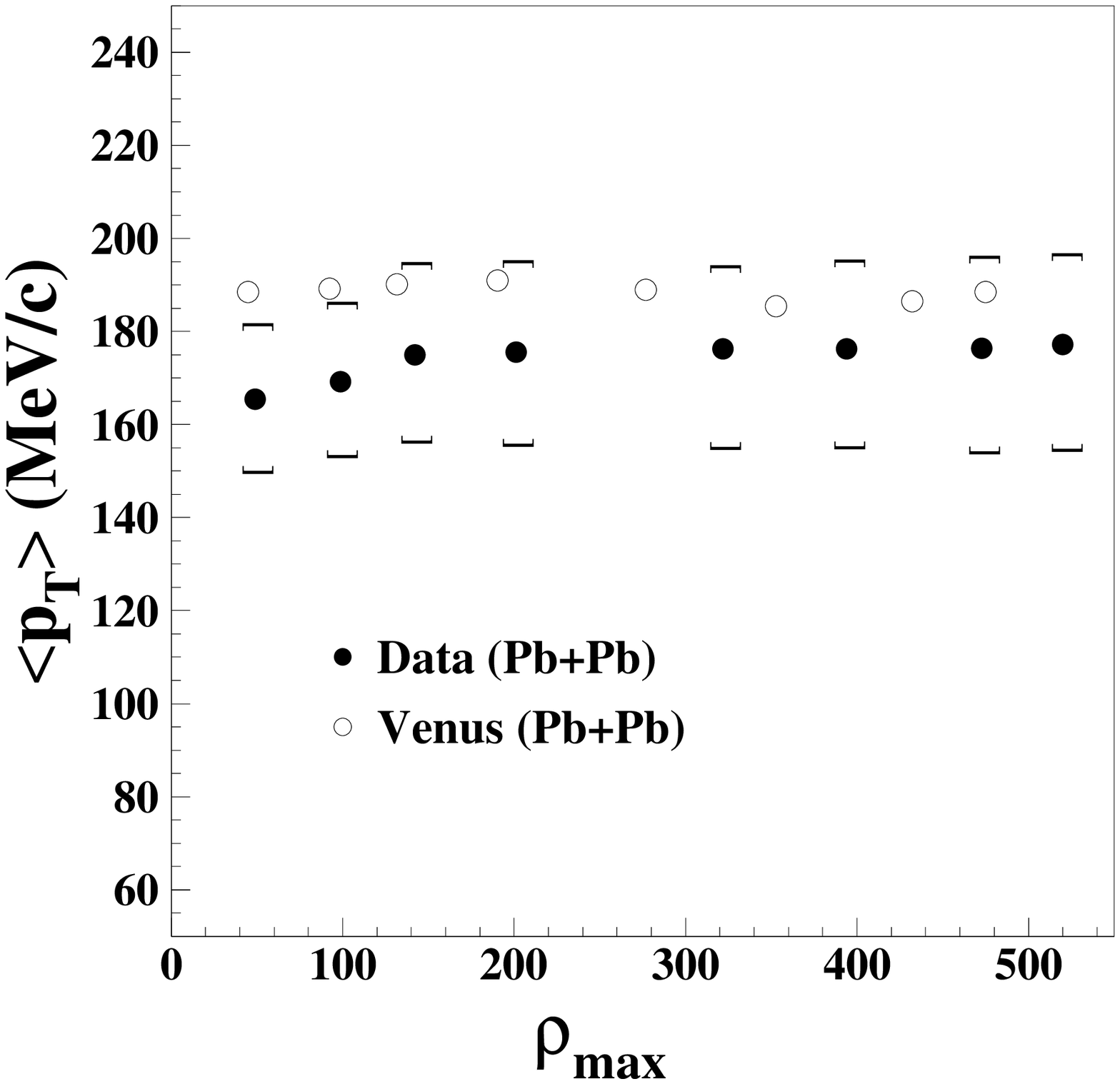}
}
\end{picture}
\vspace*{-1cm}
\caption{The mean transverse momentum, $<p_{T}>$, of
  photons as a function of centrality. For the S+Au collisions (left
  plot) centrality is defined by $E_{ZDC}/E_{beam}$ for the Pb+Pb
  collisions it is defined through the pseudo-rapidity density of
  photons at mid-rapidity,$\rho_{max}$. The $<p_{T}>$
  values obtained from the VENUS event generator are superimposed for
  comparison.
}
\label{meanpt}
\end{figure}

We have measured event-by-event $<p_{T}>$ of photons in
both WA93~\cite{pt_wa93pmd} and WA98 experiments~\cite{rap_wa98pmd}, 
using the relation  $\langle p_T\rangle
= {E^{em}_T}/{N_\gamma}$. Where $E^{em}_T$ is the transverse
component of the electromagnetic energy measured in the mid-rapidity 
calorimeter, and $N_\gamma$ is the number 
of photons measured by PMD in a given common $\eta$-$\phi$ region 
of the two detectors. The results are shown in the
Figure~\ref{meanpt}. One observes that for both WA93 and WA98 
experiment, the $<p_{T}>$ values are constant as a
function of centrality of the reaction within the
quoted systematic errors and agrees well with those
obtained from VENUS. 

\subsection{Event-by-event fluctuation in photon multiplicity}

\begin{figure}
\setlength{\unitlength}{1.1mm}
\vspace*{1.5cm}
\begin{picture}(130,50)
\put(10,5){
\epsfxsize=7.1cm
\epsfbox{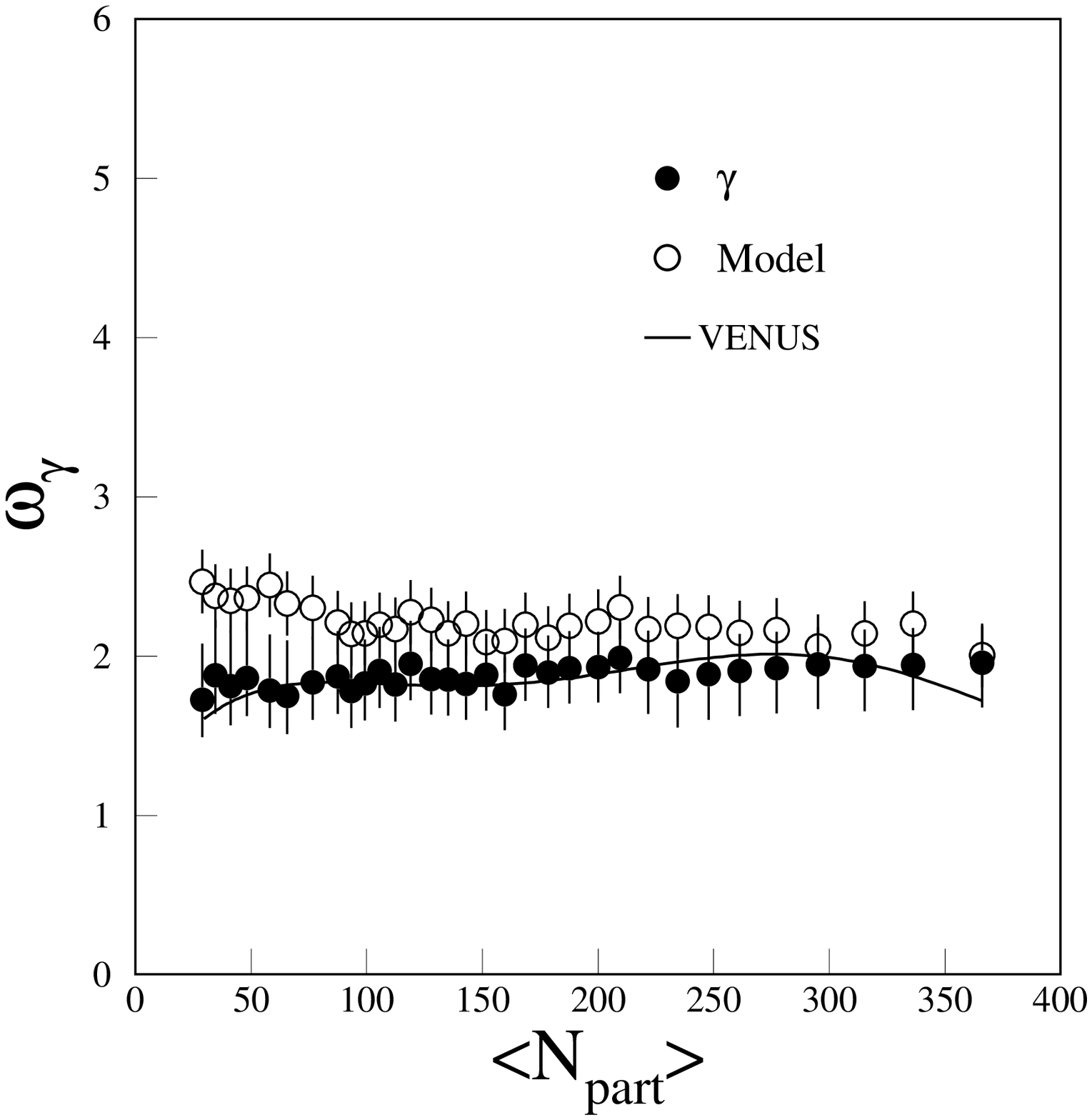}
}
\put(70,5){
\epsfxsize=7.1cm
\epsfbox{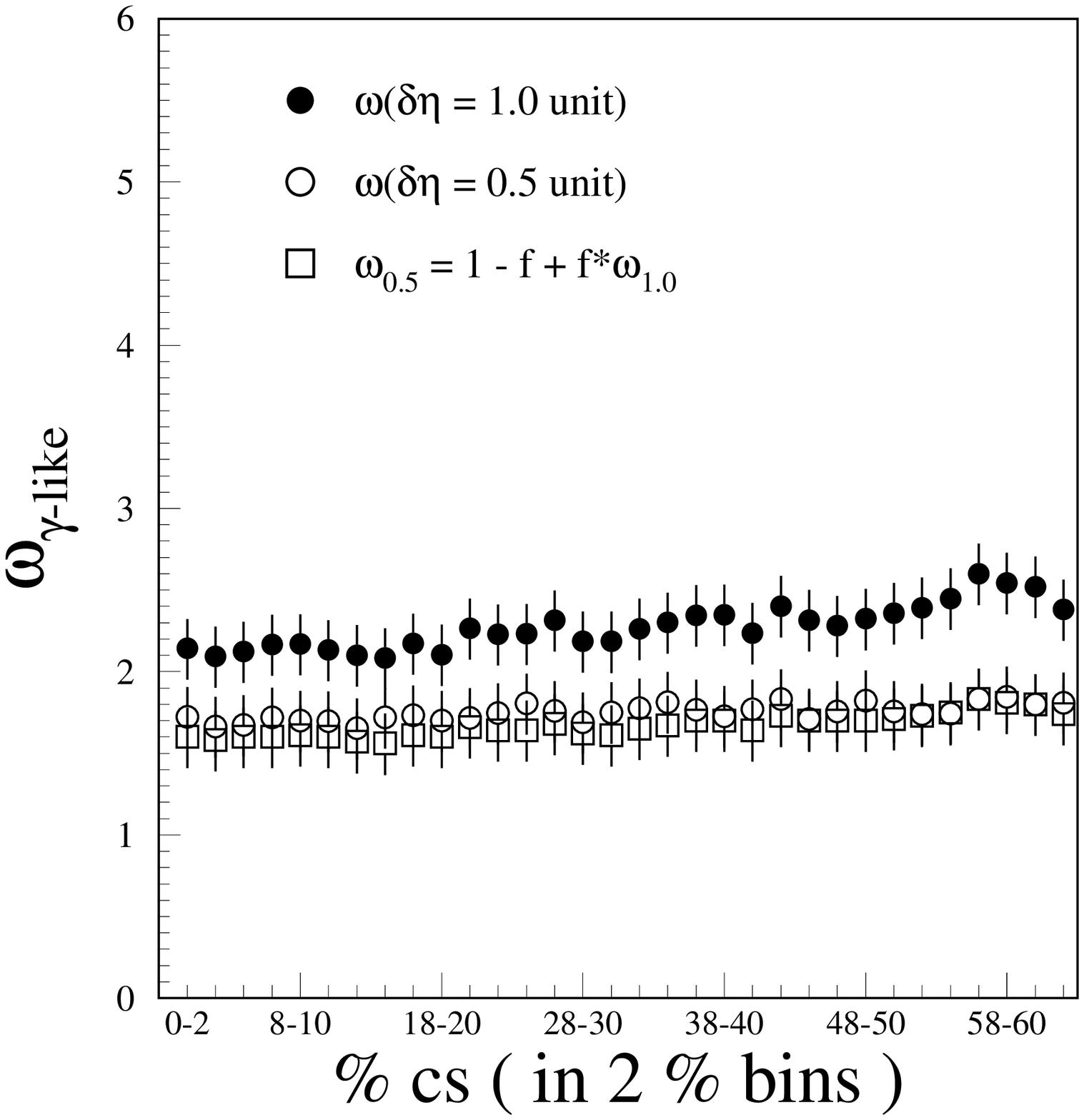}
}
\end{picture}
%\vspace*{-1cm}
\caption{
Left plot : Relative fluctuations, $\omega_{\gamma}$ of photons as a
function of number of participants. These are compared to calculations 
from a participant model and those from VENUS event generator.
Right plot : Photon multiplicity fluctuations for two $\eta$
acceptance selections. The open squares  represent estimated values of 
fluctuations in 0.5 unit of $\Delta \eta$ from the observed
fluctuations in 1.0 unit $\Delta \eta$. $f$ is the ratio of mean
photon multiplicity for 0.5 unit of $\Delta \eta$ to that for 1.0 unit 
of $\Delta \eta$.
}
\label{fluc}
\end{figure}

Event-by-event fluctuations in global observables such as photon
multiplicity has generated lot of interest in recent
years~\cite{henning}. This
is because fluctuations in the global observables are related to
thermodynamical properties of matter, more specifically multiplicity
fluctuations can be related to matter compressibility. Further studying 
the centrality and rapidity acceptance of fluctuations can help in
finding the possible existence of the $tri-critical$ point in the QCD
phase diagram~\cite{stephanov}. 
We have studied the centrality dependence and rapidity
dependence of fluctuations in photons and the results are shown in
Figure~\ref{fluc}~\cite{fluctuation_wa98}. 
The fluctuation in photon multiplicity ($\omega_{\gamma}$) 
is defined by the relation  $\omega_{\gamma} =
{\sigma_{\gamma}}^{2}/<N_{\gamma}>$. Where $\sigma_{\gamma}$ is the
standard deviation and $<N_{\gamma}>$ the mean of the gaussian photon
multiplicity distribution for a particular centrality class.
For results on 
the centrality dependence of photon multiplicity 
fluctuation the data has been compared to the results obtained from 
VENUS event generator and a simple participant model. The results of
fluctuation in photon data are
presented as a function of mean number of participant nucleon and are found 
to be in agreement with those form the model calculations within the
quoted systematic errors. For the rapidity acceptance studies, the
decrease in fluctuation with decrease in acceptance has been explained 
using a simple statistical model. The model assumes that the 
distribution  of particles in a smaller rapidity window $\Delta \eta$ 
follows a binomial sampling~\cite{accp_fluc}. 
The results have been presented as a function
of $2\%$ bins in cross section of total transverse energy.

\subsection{Formation of Disoriented chiral condensates at SPS}

  The formation of hot and dense matter in high energy heavy-ion collisions
  has the possibility of creating a chiral symmetry restored phase in the
  laboratory. A consequence of this is the possibility of
  the formation of  disoriented chiral condensate (DCC)\cite{raj}.
  The detection and study of DCC is expected to provide valuable
  information about the chiral phase transition and vacuum structure of strong
  interactions.

  DCC formation in a given domain would be associated with large
  event-by-event fluctuation in the ratio of neutral to charged pions
  in a certain domain. Since neutral pions decays to photons and
  majority of the charged particles are charged pions. The
  experimental signature of formation of DCC would be large
  event-by-event localised fluctuation in ratio of photons to charged
  particles. Following this theoretical predictions,
  WA98 recently carried out a detailed study of central Pb+Pb events
  and has been published in Ref. ~\cite{dcc_wa98}.

   The data presented here were taken during December 1996 at CERN
   SPS with 158$\cdot A$~GeV Pb ion beam on a Pb target of thickness
   213~$\mu$m. The Goliath magnet was kept off during these runs.
   A total of 85K  central events, corresponding to the top $5\%$ 
   of the minimum bias
   cross section as determined from the total transverse energy measured
   by the mid-rapidity calorimeter (MIRAC), are analyzed.
   The analyzed data corresponds to
   a common $\eta$ ($2.9< \eta < 3.75$) and $\phi$ coverage of
   charged particle and photon multiplicity detectors and for different
   segments in azimuth ($\phi$).

    The measured results are interpreted by comparison with simulated
    events and with several types of mixed events.  Simulated events
    were generated using the VENUS 4.12  event generator
    with default parameters. The output was processed through a WA98
    detector simulation package in the GEANT 3.21 
    framework. Due to the inherent uncertainties in the
description of ``normal'' physics and detector response in the VENUS+GEANT
simulations, the observation of an experimental result which
differs from the case with zero DCC fraction in simulation 
cannot be taken alone as
evidence of DCC observation.  For this reason four different types of
mixed events have been created from the real events in
order to search for non-statistical fluctuations by removing various
correlations in a controlled manner while preserving the
characteristics of the measured distributions as accurately as
possible.  The first type of mixed events (M1), are generated by
mixing hits in both the PMD and SPMD separately, with no two hits
taken from the same event. Hits within a detector in the mixed events
are not allowed to lie within the two track resolution of that
detector.  The second kind of mixed events (M2) are generated by
mixing the unaltered PMD hits of one event with the unaltered SPMD
hits of a different event. Intermediate between the M1 and M2 kinds of
mixed events is the case where the hits within the PMD are unaltered
while the SPMD hits are mixed (M3-$\gamma$), or the SPMD hits are
unaltered while the PMD hits are mixed (M3-$\mathrm ch$).
In each type of mixed event the global (bin 1) $N_{\gamma-{\mathrm
like}}$--$N_{\mathrm ch}$ correlation is maintained as in the real
event. The specific physics issues probed by each mixed event are
tabulated in Table~2.

\begin{table}
\caption{\label{tab:table1} Type of fluctuations preserved by
various mixed events}
\begin{tabular}{cccccc}
Fluctuation& &  &Mixed & Event & \\
& &M1&M2&M3-$\mathrm ch$& M3-$\gamma$ \\
\hline
$N_{\gamma}$-only& &No & Yes & Yes &No \\
$N_{\mathrm ch}-only$& &No & Yes & No &Yes \\
correlated $N_{\gamma}$-$N_{\mathrm ch} $& &No & No & No &No \\
\end{tabular}
\end{table}

\begin{figure}
\setlength{\unitlength}{1.1mm}
\vspace*{1.5cm}
\begin{picture}(130,50)
\put(10,5){
\epsfxsize=7.1cm
\epsfbox{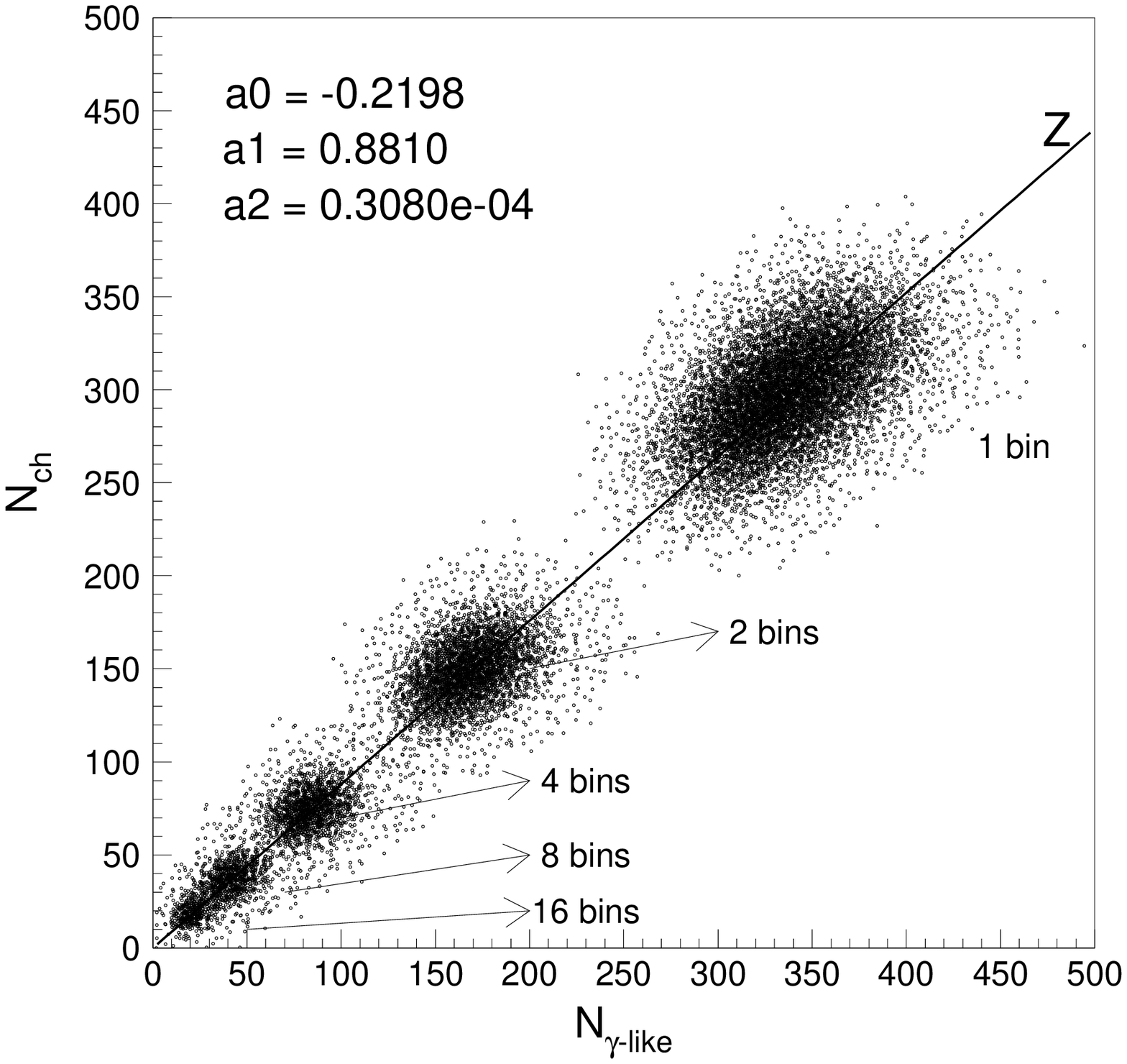}
}
\put(70,5){
\epsfxsize=7.5cm
\epsfbox{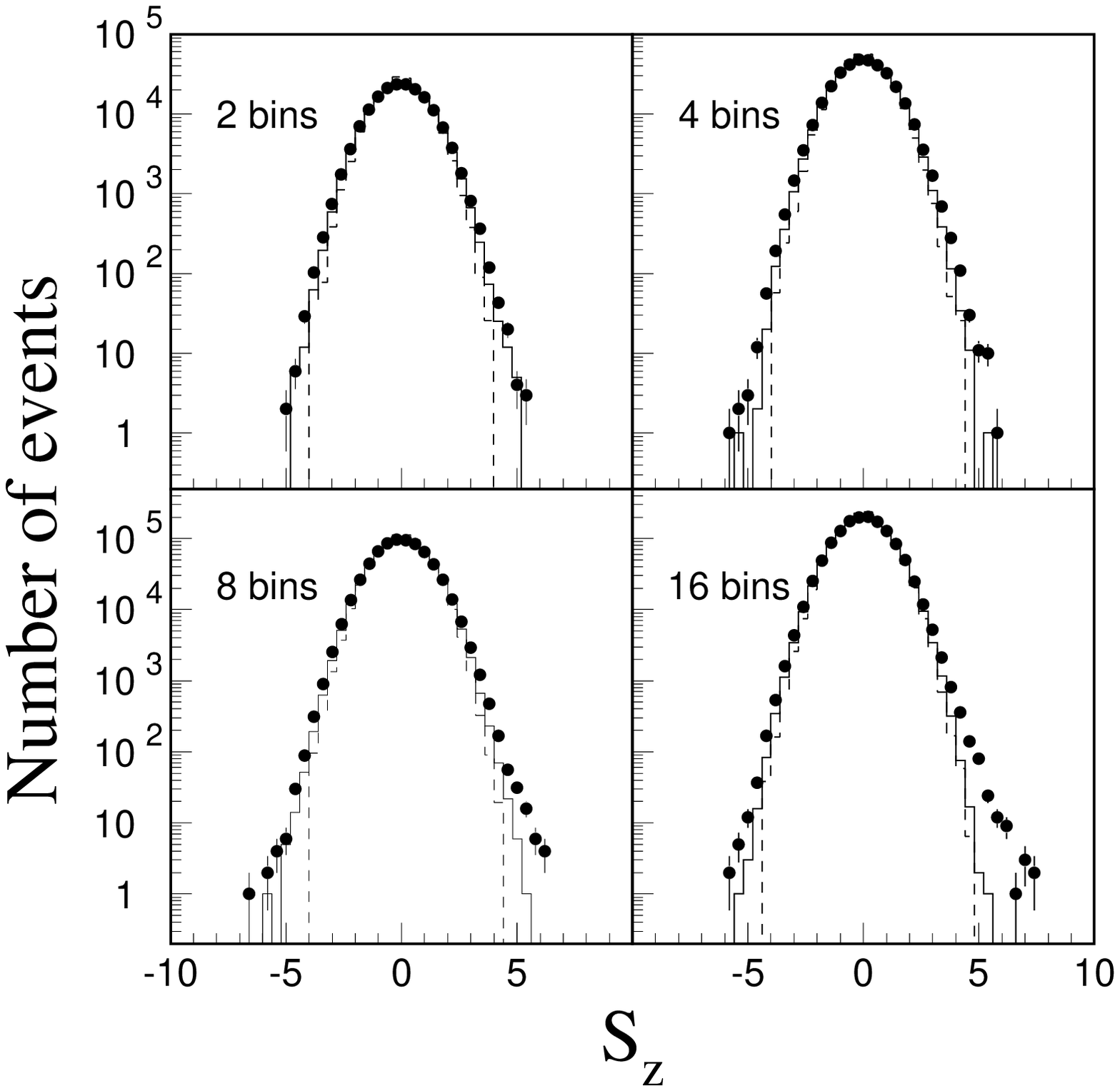}
}
\end{picture}
\caption{
Left plot : Scatter plot showing the correlation of
$N_{\gamma-{\mathrm like}}$ and $N_{\mathrm ch}$ for 1,2,4,8 and 16
bins in azimuthal angle, $\phi$, for the common coverage of the PMD
and SPMD. The common correlation axis (Z) obtained by fitting the
distributions to a second order polynomial is also shown along with
the fit parameters. Right plot : The $S_Z$ distributions for data
(solid circles), M1 
mixed events (solid histogram) and simulated events (dashed histogram)
for various bins in azimuthal angle. Distributions of other mixed
events are not shown for clarity of presentation.
}
\label{dcc}
\end{figure}

   In order to look for any possible localized fluctuation
   in photon and charged particle multiplicities, which may have
   non-statistical origin, it is necessary to
   look at the correlation between
   $N_{\gamma-{\mathrm like}}$ and $N_{\mathrm ch}$
   at various segmentation in $\eta$ and $\phi$. 
   Event-by-event correlation between $N_{\gamma-{\mathrm like}}$ and
   $N_{\mathrm ch}$ has been studied in $\phi$-segments, by dividing
   the entire  $\phi$-space into 2, 4, 8 and 16 bins.
   Fig.~\ref{dcc} shows the scatter plots of
   $N_{\gamma-{\mathrm like}}$ and $N_{\mathrm ch}$ distributions.
   The correlation plot for each of the
   $\phi$ segments,starting with the case of 1 bin which corresponds
   to no segmentation are also shown.
   A common correlation axis ($Z$) has been obtained for the full
   distribution by fitting the  $N_{\gamma-{\mathrm like}}$ and
   $N_{\mathrm ch}$ correlation with a second order polynomial.
   The correlation axis with fit parameters are shown in the figures.
   The distances of separation ($D_{Z}$) between the data points
   and the correlation axis have been calculated with
   the convention that $D_{Z}$ is positive for points below the $Z$-axis.
   The distribution of $D_Z$ represents the relative fluctuations of
   $N_{\gamma-{\mathrm like}}$ and $N_{\mathrm ch}$ from the
   correlation axis at any given $\phi$ bin.
   In order to compare the fluctuations for
   different $\phi$ bins in the same level, we use a scaled
   variable, $S_{Z} = D_Z/s(D_Z)$, where $s(D_Z)$ represents
   the rms deviation of the $D_{Z}$ distribution for VENUS events,
   obtained in a similar manner. The presence of events with localized
   fluctuations in $N_{\gamma-{\mathrm like}}$ and $N_{\mathrm ch}$,
   at a given $\phi$ bin, is expected to result in a broader
   distribution of $S_Z$ compared to those for normal events at that
   particular bin. For comparison, $S_Z$ distributions are also
   obtained for mixed events.

\begin{figure}[htbp]
\epsfxsize=8cm
\centerline{\epsfbox{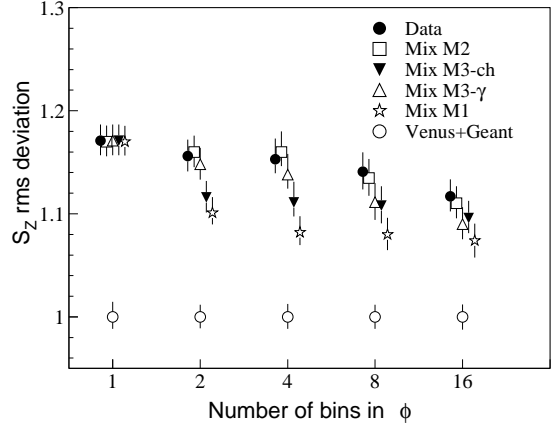}}
\caption{
The root mean square (rms) deviations 
of the $S_Z$ distributions for data, mixed events and simulated events 
as a function of number of bins in azimuthal angle.
 }
\label{sz_rms}
\end{figure}

   The $S_{Z}$ distributions  calculated at different
   $\phi$ bins are shown in Fig.~\ref{dcc} for data, M1,
   and VENUS events. The distributions for other mixed events
   are not shown for clarity of presentation. The small differences
   in the $S_Z$ distributions have been quantified in terms of the
   corresponding rms deviations  of these distributions
   as shown in Fig.~\ref{sz_rms}. The statistical errors on the values
   are small and are within the sizes of the symbol.
   The bars represent statistical and systematic errors added in
   quadrature. 
   The rms deviations of M2-type of mixed events are found
   to agree with those of the experimental data within errors.
   Thereby suggesting the absence of
   event-by-event localized correlated fluctuations in
   $N_{\gamma-{\mathrm like}}$ and $N_{\mathrm ch}$ or DCC-type
   fluctuations. The rms deviations of M1-type of mixed events are 
   found to be lower than those
   obtained for data for 2, 4 and 8 bins in $\phi$.
   The results form M3-type of mixed events are found
   in between those obtained form data and M1-type mixed events. 

   These
   results indicate the presence of localized fluctuations in data, due to
   both photons and charged particles for certain  bins in $\phi$.
   Within the
   context of a simple DCC model~\cite{dcc_wa98}, 
   upper limits on the presence of
   localized non-statistical DCC-like fluctuations has been set to be 
   $10^{-2}$ for DCC domain size with 
   $\Delta\phi$ between 45--90$^\circ$ and $3\times 10^{-3}$ for
   $\Delta\phi$ between 90--135$^\circ$ with $90\%$ confidence level.

\section{Photon multiplicity detectors at RHIC and LHC}

Photon multiplicity detectors will be installed in STAR experiment at
RHIC and in ALICE at LHC. The detector technology is different from
those used at SPS. PMD in STAR and ALICE will be a gas based detector
with honeycomb cells~\cite{alice_pmd}. 
Typical detector parameters for the PMD to be used
in STAR and ALICE is given in the Table~3. 

\begin{table}[h]
\caption{PMD in STAR and ALICE }
\begin{tabular}{lll}
Basic features & STAR-PMD & ALICE-PMD \\ \hline
Expected year of Installation & 2002 & 2005 - 2006 \\ 
Beam and target & Au + Au & Pb + Pb \\ 
CMS Energy & 200 A GeV & 5.5 A TeV\\ 
No. channels & 82944 & 270,000 \\
Readout & Gassiplex+CRAMS & Gassiplex+CRAMS \\
Distance from target & 5.5 m & 3.5 m \\ 
$\eta$ coverage & 2.3 - 3.9 & 2.3 - 3.5 \\
Efficiency (central) & 62 \% & 64 \%\\ 
Purity (central) & 61\% & 60 \%\\ 
\end{tabular}
\end{table}

PMD will address all the  physics issues discussed in Section~3. 
It will have overlap of about one unit in $\eta$ with full $\phi$ 
with the forward time projection chamber (FTPC) in STAR. The momentum
information of charged particles in FTPC will strengthen the DCC-type
of study at RHIC. This is demonstrated by a simple calculation given
below. In addition, PMD will be able to give estimates of transverse
electromagnetic energy in ALICE. High multiplicity environment at ALICE
will be suitable for such estimates as the statistical fluctuations on 
transverse electromagnetic energy measurements gets reduced.

\subsection{DCC search in STAR}

  The $p_{T}$ information of charged particles will enhance the 
  possibility of DCC search and verify the various features of
  DCC domain, such as DCC-pions are low $p_{T}$ pions and DCC formation
  may lead to low $p_{T}$ enhancement in pions. The photon multiplicity
  detector in-conjunction with a charged particle detectors which gives
  multiplicity  along with particle-wise momentum information, such as 
  the FTPC in STAR experiment can be
  used. 

\begin{figure}[htbp]
\epsfxsize=8cm
\centerline{\epsfbox{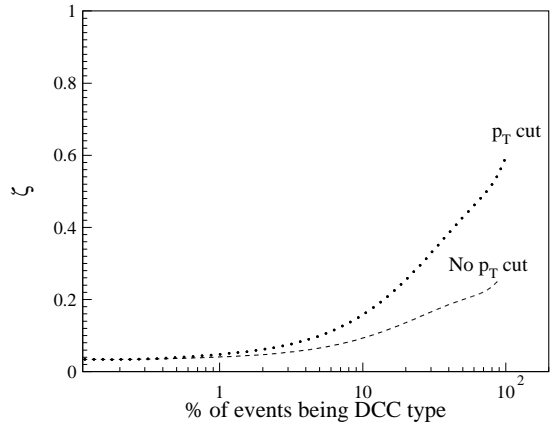}}
\caption{
Variation of $\zeta$ as a function of $\%$ of events being
  DCC-type, for charged particle detector with $p_{T}$ information.
  Also shown are the corresponding results with charged particle
  detector without $p_{T}$ information.
}
\label{lowpt}
\end{figure}

  In order to show the utility of $p_{T}$ information we carried out
  the following simulation study. The basic detector setup for DCC
  study was
  implemented by putting a photon multiplicity detector 
  and a charged particle multiplicity detector 
  with momentum information, having a common coverage (as in STAR ) with
  realistic detector parameters. The photon counting efficiency was 
  taken as $70\% \pm 5\%$ and contamination in photon sample was taken 
  to be $30\% \pm 5\%$. The detectors accepts particles with
  momentum above $30~MeV/c$. The efficiency of charged particle
  detector was taken as $90\% \pm 5\%$ with $p_{T}$ 
  resolution of $\frac{\Delta p_{T}}{p_{T}}$ = $0.2$.
  Then a DCC-type fluctuations in low $p_{T}$ pions 
  was introduced through a simple DCC
  model as described in Ref.~\cite{dcc_wa98}.  
  Events with different fractions of
  DCC-type events were generated by mixing normal events and DCC-type
  events. Then the resultant events were analysed using DWT analysis.
  The analysis was done with and without $p_{T}$ cut on charged
  particle data($p_{T}$ < 150 MeV/c). 
  The results are shown in Figure~\ref{lowpt}.
  One finds the strength, $\zeta$, of the DCC signal as a 
  function of percentage of events being DCC-type is higher with  $p_{T}$
  cut than without $p_{T}$ cut. This clearly demonstrated that 
  with $p_{T}$ information of charged particles the sensitivity
  to search for DCC at RHIC increases.

\section{Summary}

A high granularity preshower photon multiplicity detector was
successfully implemented in the WA93 and WA98 experiments at CERN SPS.
The PMD provided the $first~ever$ measurement of
multiplicity, rapidity and azimuthal distributions of photons in 
ultra-relativistic heavy-ion collisions. Analysis of these
distributions showed that photon multiplicity is an important
global observable, providing significant
information regarding particle production and reaction mechanism in
heavy-ion collisions at relativistic energies.  
The study of pseudo-rapidity distributions of photons,
event-by-event measurement of mean transverse momentum of photons,
scaling of photons and event-by-event fluctuations in photon
multiplicity suggest that it is complementary to charged particle
measurements. In addition to this, correlation between photon and 
charged particle multiplicity has been found to
provide important information regarding formation of
DCC in relativistic heavy-ion collisions. A gas based PMD will be
implemented in the STAR experiment at RHIC and in ALICE experiment at
LHC. It will address all the physics issues, which were studied using
the PMD at CERN SPS. However, DCC search will get a boost due to momentum
information of charged particles at STAR. With very high multiplicity
environment at ALICE, PMD can provide estimates of transverse
electromagnetic energy at LHC.

{\bf Acknowledgments} \\
We wish to thank all the collaborators of WA93 and WA98 experiments
for their active participation in the analysis of data from PMD, 
which resulted in many good publications, most of which is discussed 
in this work. We also wish to thank our STAR and ALICE Collaborators 
who have contributed to the proposed use of PMD in these experiments.
We acknowledge with appreciation the effort of all engineers,
technicians, and support staff who have participated in the
construction of the photon multiplicity detector and runing of the
WA93 and WA98 experiments. This work was supported jointly by the 
Department of Atomic Energy, the Department of Science and Technology, 
the Council of Scientific and Industrial Research and the University Grants 
Commission of the Government of India.

\end{document}